# UNCOVERING BIAS IN ORDER ASSIGNMENT*

Darren Grant
Department of Economics and International Business
Sam Houston State University
Huntsville, TX

dgrant@shsu.edu

Abstract: Many real-life situations require a set of items to be repeatedly placed in a random sequence. In such circumstances, it is often desirable to test whether such randomization indeed obtains, yet this problem has received very limited attention in the literature. This paper articulates the key features of this problem and presents three general tests that require no a priori information from the analyst. These methods are used to analyze the order in which lottery numbers are drawn in Powerball, the order in which contestants perform on *American Idol*, and the order of candidates on primary election ballots in Texas and West Virginia. In this last application, multiple deviations from full randomization are detected, with potentially serious political and legal consequences.

---

* I appreciate the assistance of Elsy Orellana and Jackie Sanchez in assembling data and the contributions throughout from Sheridan Grant, who scraped data, assisted with some of the estimation, and provided constructive commentary and feedback. In addition, a couple of West Virginia county clerks provided valuable insights. This project was supported by grants from Sam Houston State University's Internal Grant Program and College of Business Administration. I am willing to share all of the data used in this paper for any purpose, with proper credit given.

Individuals are forced to prioritize in many aspects of life: in judging athletic contests, voting in elections, evaluating applicants for a scholarship or a promotion, and so on. When the rankings thus produced affect others, as in these examples, fairness and efficiency dictate that we minimize any bias involved in their formation.

One such bias involves the sequence in which the individuals are presented for consideration, which will be called an *ordering*. For example, gymnastics competitions present the contestants one after another; elections present candidates for office in a vertical order on a ballot; final admissions decisions at most elite universities are made one applicant at a time.

In such situations, cognitive bias can affect the way that contestants are assessed, favoring those presented toward the beginning or end of the sequence. Several explanations for such effects have been offered. Krosnick (1991) and Miller and Krosnick (1998) hypothesize that people "tend to evaluate objects with a confirmatory bias," looking for reasons to accept rather than reject them. As they work through the list of options, mental fatigue can lend an advantage to the objects listed earlier in the sequence. Alternatively, Mussweiler (2003), Damisch, Mussweiler, and Plessner (2006), and others emphasize that the assessment of any one item in a sequence may be influenced by a comparison with its predecessor, in ways that can advantage the later items in the sequence. Salant (2011) argues that order effects are an "inevitable" consequence of "bounded rationality."

They are certainly ubiquitous in real life. A significant body of empirical evidence has arisen in three areas: contests, consumer choice, and voting. In figure skating and music competitions, the evaluations given to contestants tend to increase in sequential order, so that the first contestant tends to be judged most harshly and the last-considered judged most favorably (Bruine de Bluin, 2006; Glejser and Heyndels, 2001; Haan, Dijkstra, and Dijkstra, 2005; Page and Page, 2010; Antipov and Pokryshevskaya, 2017). On the other hand, in consumer choice, the first good offered is generally preferred (Dean, 1980; Mantonakis et al., 2009; Carney and Banaji, 2012; Novemsky and Dhar, 2005). This primacy effect occurs even more strongly in voting, where over a dozen studies find that the candidate listed first on the ballot receives a boost of 1-5% of the total vote, depending on the contest examined, a phenomenon called the "ballot order effect" (Miller and Krosnick, 1998; Meredith and Salant, 2013; Grant, 2017, and cites therein).[1]

Circumstances like these open the door to a principal-agent problem, whereby the ordering of contestants is *manipulated* to advantage preferred individuals or discriminate against ill-favored individuals, to the detriment of the contest's integrity and societal welfare. Randomly determining contestant order minimizes any such favoritism, and many entities have rules to this effect. For example, the order of countries in each year's *Eurovision* song contest is chosen by lot, while the order of candidates on each county's primary election ballot is randomly determined in several states.

In these two examples, orderings are repeatedly conducted for the same set of countries or candidates. In such situations one can test, statistically, whether they have indeed been conducted randomly. If the set of orderings is unlikely to have occurred by random chance, one can conclude that some orderings in the set have been manipulated to favor or disfavor certain contestants. For example, Grant (2017) found statistically significant deviations from randomness in multiple primary elections in Texas–a sign that some county election officials were ignoring the law. When it is impractical to monitor the determination of individual orderings, these tests would be the only way to ascertain that such conduct exists.

Such tests can be found, occasionally, in the empirical literature, but the methods used to conduct them have been improvised and ad hoc. Page and Page (2010), examining the program *American Idol*, investigate whether “strong” contestants who scored well in prior rounds are more likely to be presented at the beginning of the show or the end. Meredith and Salant’s (2013) analysis of candidate order on election ballots proceeds similarly, examining whether incumbents’ ballot positions are evenly distributed. Both approaches use limited information, as they ignore the positions of other contestants. Other approaches, though similarly ad hoc, are more inclusive. Ho and Imai (2008) examined the average difference in rank between pairs of letters in randomized alphabets used for ballot orderings in California, while Grant (2017) applied the Fisher Exact Test (later shown to be anti-conservative) to the cross-tabulation of ballot positions of candidates in primary elections. Nonetheless, these studies also use limited information, as they rely only on the aggregated tabulation of ranks across letters or candidates, not the individual orderings themselves.

[1] Sequencing effects also exist in theory. For example, in the Secretary Problem in mathematics and some search models in labor economics, job applicants that appear early in the sequence are treated differently, or held to a different standard, than those presented toward the end.

There is thus a need for the systematic development of best-practice methods to test a set of orderings for randomness. Such methods would have greater power to detect deviations from randomness in a variety of circumstances and greater credibility in situations in which failure to randomize appropriately is penalized.

An initial attempt at doing so is a recent paper by Grant, Perlman, and Grant (2020, hereafter GPG). This paper derives two methods, described below, to test for deviations from randomness that are specified by the analyst in advance. These *targeted* tests require the analyst to anticipate how any such orderings would be manipulated. In some circumstances this assumption can be reasonable, such as ordering a set of candidates on a ballot, one of which is far more popular than the others. But in most circumstances this is not so. Then these targeted tests are unsuitable. To provide a fully general solution to this general problem, we need *untargeted* tests that can detect any type of deviation from randomness, statistical power permitting. These tests would be defensible in any context, filling a major gap in the literature.

This paper introduces three such tests and examines their usefulness using both simulations and real-life applications. In suitably large samples, these tests reliably detect systematic manipulation of the orderings with no a priori information required. Such manipulation is observed in one of our applications, with potentially serious legal and political ramifications.

That is this paper's immediate purpose. But it has a larger purpose as well, which is to fully articulate the problem to begin with. Applied work in this area has many facets: computational issues, the structure of the data, power and robustness considerations, the motivations for deviating from randomness and the nature of the preferences thereby indulged. As this literature is embryonic, most of these have not been previously examined or even articulated as issues; the others have been examined by GPG in a limited fashion suited to the targeted tests developed in that paper. Here we explore them all in depth, delineating the problem in its fullness and laying out the landscape of practical issues empirical analysts can face in testing a set of orderings for randomness.

Accordingly, we analyze an eclectic mix of real-life applications: the sequence in which contestants perform on the popular television show *American Idol*; the order in which the "white balls" are drawn in the largest lottery in the U.S., Powerball; and the order in which candidates are listed on election ballots in Texas and West Virginia. These applications have varied types and strengths of preferences; varied data structures and statistical power; mechanical, pro-social, and anti-social reasons for manipulating the orderings; and wide-ranging computational demands. The outcomes are equally varied: deviations from randomness take a variety of forms, their differences generated by seemingly minor institutional details.

Altogether, this variety drives this paper's ultimate conclusion: there is no single "best test" for the randomness of a set of orderings. The optimal test depends on the circumstances, and can include a test that was developed previously, one of the tests put forward in this paper, or an adaptation of one of these tests. To be fully equipped for empirical work in this area, researchers must understand the landscape of the problem and have this battery of tests at their disposal. Only then can they choose an analytical method that adequately suits the purpose.

## I. Three New Tests–and Three Old Ones.

Setup. Following GPG, let there be K *items*, indexed by k = 1..K, which are ordered by each of N *agents*, indexed by i = 1..N. (Alternatively, a given agent could order these items N times.) An ordering is an arrangement or permutation $\Pi \equiv (\pi_1, \pi_2 .. \pi_K)$ of these items such that item k is assigned *position*, or rank, $\pi_k$. Naturally, these orderings can be indexed by i, $\Pi_1, \Pi_2 .. \Pi_N$.

In "ranking models," all agents are intended to rank these items purposively. In doing so they are (imperfectly) guided by an objective *preference criterion*, which these models then estimate (see, for example, Fligner and Verducci, 1986). We posit the obverse: all agents are expected to randomly assign items to positions, so that Π is uniformly randomly distributed and exchangeable. This serves as our null hypothesis. The alternative is that a non-empty subset of agents manipulate their orderings in accordance with their own subjective preferences.[2] In the targeted tests presented by GPG, these choices are assumed to be governed by a preference criterion that is known by the analyst a priori. The untargeted tests introduced here do not make this assumption. The null hypothesis can be rejected for any sufficient deviation from randomness, of any type, which is then taken to be evidence of manipulation of the orderings by a subset of agents. Section II will describe some different types of preference criteria and indicate which tests best suit each.

[2] Implicitly, these preferences factor in any order effects on decision-making. Thus, when there are primacy (recency) effects, the agent would tend to list their preferred items first (last). This point is inconsequential for the purposes of this section.

We discuss and ultimately compare all new or existing tests that examine all items contained in the ordering. These tests fall into three groups: the targeted tests developed by GPG, *aggregated* untargeted tests that rely only the cross-tabulation of items and their positions, and *disaggregated*, micro-level untargeted tests that utilize the individual orderings.

Targeted Tests. The two targeted tests developed by GPG are the Rank Compatibility Test and the Linear Concordance Test.

**The Rank Compatibility Test.** This test determines whether orderings that are *fully compatible* with posited preferences, ties being allowed, are unusually frequent. For example, let preferences over items A-F be A ~ B ≻ C ~ D ~ E ≻ F, where ≻ indicates preference and ~ indifference. Then the ordering {A, B, E, D, C, F} would be fully compatible, while the ordering {A, E, B, D, C, F} would not, since B ≻ E. Given any preference criterion, the number of fully compatible orderings takes a binomial distribution under the null. Using this fact, GPG's test determines whether the observed frequency of such orderings is unusually large.

**The Linear Concordance (LC) Test.** This test is sensitive not only to orderings that are fully compatible with posited preferences, but also to those that are *partly compatible*, such as {A, E, B, D, C, F} in the example above. As GPG note, such orderings would occur if the positions were filled sequentially and "each item's selection probability is an increasing function of perceived social status or a decreasing function of some discriminatory trait. [They] could also occur because of minor differences in biased agents' preferences...or other small frictions in the generation of a biased ordering from a preference criterion" (p. 6). Such variation is also posited by ranking models (e.g., Mallows, 1957).

To execute this test, agents' preferences are represented by a score vector, $s \equiv (s_1, s_2, .. s_K)$, that is prescribed in advance. One then computes the mean concordance of this vector with the positions of each item in a set of orderings $\Pi_i$, i = 1..N. For *admissible* scores that have a mean of zero and a sum of squares of one, GPG show that this product, suitably scaled, asymptotically has a standard normal distribution under the null:

$$L = \sum_{i=1}^{N} \sum_{k=1}^{K} s_k \pi_{i,k} / \sqrt{NK(K+1)/12} \; \overset{a}{\sim} \; N(0,1) \qquad \textbf{(1)}$$

where L is the scaled linear concordance. The score vector can reflect the presumed strength of preferences or merely their ordinal ranking (with ties permitted). In the latter case, GPG show that the optimal scores are the scaled, demeaned ranks (averaged over ties).

When the analyst correctly anticipates the preference criterion, these two tests should supersede untargeted tests that do not utilize this a priori information. Without this information, however, these targeted tests are impotent. Untargeted tests are needed instead.

<u>Aggregate Untargeted Tests</u>. There are also two of these: the Max LC Test, which is new to the literature, and Ho and Imai's Rank Test.

**The Max LC Test.** This transforms the LC Test into an untargeted test. Instead of prescribing a score vector in advance, it searches for the *feasible* score vector that yields the largest value of L, called $L^*$.

There are two ways that "feasible" can be defined in this context. The first presumes only an ordinal ranking of items, remaining agnostic on the strength of those preferences, as in the A-F example above. For each possible preference ranking, the optimal score vector is determined as described above and the value of L determined. Selecting the largest of these across all possible preference rankings yields the $L^*$ associated with a given set of orderings.

In this *strictly untargeted* approach, the set of feasible score vectors is finite, limited to the size of the set of possible preference rankings. This set grows rapidly in K.[3] Including ties, there are 75 possible preference rankings when K=4 and 4,683 when K=6. In this latter case, the set of feasible scores forms a fine grid spread evenly throughout the admissible portion of "score space."

In such cases it is just as well and far simpler computationally to ignore the discretization, allow all admissible scores to be feasible, and solve directly for the optimal score vector $s^*$. This approach, called *freely untargeted*, also applies when one is willing to quantify the strength of preferences. The program that determines $s^*$ is solved in Appendix A, yielding:

$$s_k^* \propto \sum_{p=1}^{K} pC_{k,p} - N(K+1)/2 \quad \textbf{(2)}$$

where $C_{k,p}$ is the number of times item k has been placed in the $p^{th}$ position. One then normalizes the scores to have a 2-norm of one and calculates the associated value of $L^*$. In either case it is straightforward to determine the distribution of $L^*$ under the null using Monte Carlo simulation.

---

[3] The set has $\sum_{k=1}^{K} k!S(K,k)$ elements, where S is the Stirling number of the second kind.

**The Rank Test.** This test is based on the mean absolute difference in ranks between any two items. The test statistic R is calculated as follows:

$$R = \frac{\sum_{j \neq j^*} \frac{1}{N} \left| \sum_{i=1}^{N} (\pi_{ij} - \pi_{ij^*}) \right|}{K(K+1)/2} \qquad \textbf{(3)}$$

where $\pi_{ij}$ is the position, or rank, of item j in ordering i. The p-value is determined through Monte Carlo simulation.

This test is closely connected to the Max LC Test, because, per eq. **(2)**, the optimal score vector in the freely untargeted version of that test consists of the scaled mean ranks. To get the associated value of $L^*$, this score vector is multiplied by the ranks themselves, which means that $L^*$ is a function of the squared mean ranks. We should expect similar p-values across the two tests.

Disaggregated Untargeted Tests. There are two of these, both new to the literature: the Equality of Permutations Test and the Cascading Chi-Squared Test.

**Equality of Permutations Test.** This test examines whether each permutation in the sample occurs with equal likelihood. When K = 2, there are only two possible permutations, so this test devolves into the binomial test of equal proportions. For K ≥ 3, we use the basic chi-squared test of equal proportions:

$$t^E = \frac{\sum_{p=1}^{K!} (C_p - N/K!)^2}{N/K!} \sim \chi^2(K!-1) \qquad \textbf{(4)}$$

where p = 1..K! indexes the set of all possible permutations and $C_p$ is the count of observed occurrences of permutation p in the sample.

This test is simple, straightforward, and flexible. However, the number of permutations grows quickly in K, mandating a large sample size, N, in order to satisfy the standard requirement for a chi-squared test that all expected frequencies be at least 5, i.e., $N/K! \geq 5$. Size calculations let us relax this rule somewhat: for $K \geq 4$, a guideline of $N/K! \geq 2$ works well in practice, perhaps because of the large number of permutations that are checked.[4] Nonetheless, even this looser guideline requires well over a thousand orderings when K is only 6.

**The Cascading Chi-Squared Test.** This test generalizes the well-known chi-squared test of equal proportions so that it can be applied to orderings (though in a different way than above).

Counting the frequency that a given item, item k, is placed in 1st position, 2nd position, etc., in a set of orderings, one can calculate the appropriate chi-squared statistic, namely:

$$t^k = \frac{\sum_{p=1}^{K} (C_{k,p} - N/K)^2}{N/K} \sim \chi^2(K-1) \tag{5}$$

where p = 1..K indexes positions and $C_{k,p}$ is the count of observed occurrences of item k in position p. One could do this for all items k = 1..K. However, the position of one item in an ordering affects the position of the others, so these placements are not independent. One could calculate $t^k$ for each item, but they could not be meaningfully combined.

Our response is to adapt this approach so that the chi-squared statistics are independent. After calculating the chi-squared test statistic for one item in a set of orderings, simply "remove" that item from the set and analyze the order of the remaining items only. For example, in the A-F example above, count the frequency that A is placed in 1st position, 2nd position, etc., and calculate $t^A$, which is distributed $\chi^2(5)$ under the null. Then remove item A from all orderings, so that {A, E, B, D, C, F} becomes {E, B, D, C, F} and so on, and repeat the computation for item B. Under the null, that test statistic, $t^B$, is distributed $\chi^2(4)$ and is independent of its predecessor. Continuing this process to completion and summing the $t^k$'s that result yields a single test statistic for the randomness of the overall ordering, which is asymptotically distributed $\chi^2(K(K-1)/2)$.[5]

[4] When N = 2K!, simulated size at $\alpha$ = .01, .05, .10 is {.008, .034, .089} for K = 3, {.012, .045, .086} for K = 4, {.010, .050, .095} for K = 5, and {.009, .049, .099} for K = 6.

## II. Simulations.

Setup. We determine the power of each test as a function of the number of items being ordered, K; the number of orderings, N; and the "strength" of preferences. We permit orderings that loosely correlate with preferences. That is, each ordering need not be fully compatible with the preference criterion, though the average position of each item in a large set of orderings will so conform.

Therefore, our simulations generate each ordering sequentially, selecting the next item from the remaining items with relative probabilities set as follows: $P_{k+m} / P_k = \delta^m$, $\delta \in (0,1]$, where k and m are positive integers, $k+m \leq K$, and $P_j$ is the probability of selecting item j. The term $\delta$ represents the relative probability of item j being chosen relative to item j-1. When $\delta = 1$, there are no preferences and each ordering is randomly determined. When $\delta < 1$, item 1 $\succ$ item 2 $\succ$ item 3, and so on, with these preferences becoming stronger as $\delta$ falls. Targeted tests anticipate that this preference criterion governs any deviations from randomness, while untargeted tests do not.

[5] The final test statistic is not impervious to the order in which the items are selected for analysis. Thus, proceeding through items A-F in, say, reverse alphabetical order will not generally yield an identical result. For this reason, it is important to proceed through the items in a pre-ordained order, such as alphabetical order or numerical order, so as to prevent p-hacking. This approach is used in the applications below.

Table B1 in Appendix B presents cross-tabulations for 10,000 orderings when $K = 4$, using the four values of $\delta$ that are employed in our simulations: 0.95, 0.9, 0.8, and 0.7. It is apparent that $\delta = 0.95$ represents weak preferences, in which the first item appears in first position just a bit more than its confederates, while $\delta = 0.7$ represents preferences that are very strong. Thus our simulations encompass the full range of preference strengths.

For K and N, we use a range of values intended to represent the wide variety of sets of orderings to be found in real life. When $K = 2$, the binomial test of equal proportions applies and the tests in this paper are unnecessary, so we choose $K \in \{3, 4, 5, 6, 10, 15, 20\}$ and $N \in \{50, 125, 250, 1000\}$. The values of K and N fall into these ranges for several of the applications presented below. (In a few cases, computational costs are $O(K!)$ and quickly explode, so only the smaller values of K are used.)

Tables 1-4 present power at $\alpha = 0.05$, using 10,000 simulations, for all six tests considered in this paper, grouping the untargeted Max LC and Equality of Permutations Tests with their targeted equivalents, the LC and Rank Compatibility Tests. Overall, rejection rates range from 0.05 to 1, increasing as K and N grow and $\delta$ falls, as expected. Accordingly, when power equals one for a given combination of K, N, and $\delta$, the entries for larger values of K and N and smaller values of $\delta$ are omitted, as they also implicitly equal one.

Results. We begin with the Rank Test, found in Table 1. This test exhibits very low power when preferences are mild and K and N are reasonably small. However, power grows steadily in both K and N. With few items, one can expect to reject the null hypothesis in small samples (N=50) when preferences are strong ($\delta \leq 0.8$) and in larger samples when preferences are weak. With many items ($K \geq 10$) rejection is likely even with small samples and weak preferences.

Next consider the Max LC test, presented in Table 2. Power for the freely untargeted version of this test mirrors that of the Rank Test, as expected, since both tests are based on mean ranks. Rejection rates are also similar for the strictly untargeted test, which is far more cumbersome to execute. However, these tests' power pales in comparison to that of GPG's targeted LC test, presented in the last panel of the table. As in GPG, knowing the preference criterion that agents use in manipulating orderings is very helpful.

We now shift from aggregated tests to disaggregated tests. Results for the first of these, the Cascading Chi-Squared Test, are found in Table 3. There is a modest reduction in power relative to the two preceding tests, but the test can still detect deviations from randomness in small samples with strong preferences or large samples with weak preferences.

Finally, in Table 4, we consider the most general test of all, the Equality of Permutations Test (capping K at five, so as to satisfy the requirement that $N/K! \geq 2$). The untargeted version modestly retreats from the power of the Cascading Chi-Squared Test, though it still retains the ability to detect deviations from randomness when samples are large or preferences are strong. Surprisingly, the targeted version, which reverts to GPG's Rank Compatibility Test, generally performs worse. This test puts all its eggs in one basket, so to speak, examining the frequency of just that single permutation that is fully compatible with the preference criterion. The untargeted test, in contrast, picks up the disproportionate number of orderings that loosely, but not perfectly, reflect preferences. This is often superior.

Power and Preferences. The preference criterion adopted in our simulations (when $\delta < 1$) is what GPG calls *unidirectional*. All agents who manipulate the orderings do so in accordance with the same hierarchical preferences, with only "random" or unsystematic deviations permitted. However, other possibilities exist.

This is most easily seen in the political context. There, unidirectional preferences would obtain if one candidate, say an incumbent, was generally more popular or highly regarded than his or her opponents, and orderings were influenced accordingly. An alternative would be *bidirectional* preferences that were based on ideology. Any agent conducting an ordering could prefer either end of an ideological spectrum to the other, for example, may prefer liberals to conservatives or vice versa.[6] Another, more complex alternative is *multidirectional preferences*, in which agents themselves are arranged on an ideological spectrum and prefer candidates with similar ideologies. In this case distinct, somewhat idiosyncratic sets of orderings will occur disproportionately, with some agents placing liberal candidates more highly, others placing conservatives highly, and others favoring moderates.[7]

The three types of preference criteria have different rank effects. The strongest occur with unidirectional preferences, which directly affect mean ranks. The weakest occur with bidirectional preferences, in which opposing agents' manipulations offset. In between lie multidirectional preferences. While opposing agents' preferences will tend to offset, moderates will still have higher ranks overall, as they are least preferred by no one.

---

[6] Within the Republican party, ideology could manifest along a spectrum of Tea Party Republicans to "establishment" Republicans; within the Democratic party, it could manifest along a spectrum of Progressives to Centrists.

[7] To be more specific, orderings that "hopscotch" across ideological neighbors are less likely, because no agent should "rationally" create them, no matter what their ideology is. For example, let candidates be arranged as follows: $A \sim B \succ C \sim D \sim E \succ F$, where $\succ$ means "more conservative than" and $\sim$ means ideological equivalence. No agent would intentionally construct the ordering {B, E, F, A, D, C}, so this ordering should be relatively infrequent.

The relative power of tests that emphasize mean ranks, the Rank Test and Max LC Test, varies accordingly. It is strongest for unidirectional preferences, as shown by our simulations, and weakest for bidirectional preferences, where the Cascading Chi-Squared and Equality of Permutation Tests dominate. Though these latter two tests have less power under unidirectional preferences, they better maintain power under other preference types.

To broaden our simulations accordingly, Appendix C generalizes the process generating orderings to incorporate bidirectional and multidirectional preferences, and presents power for these preference types for all four untargeted tests. These results affirm the conclusions asserted above. The Equality of Permutations Test is most robust under the limited circumstances that satisfy the requirement of $N/K! \geq 2$; otherwise, that appellation falls to the Cascading Chi-Squared Test. Both surpass the Rank and Max LC Tests for bidirectional preferences, while the results for multidirectional preferences are more ambiguous. The ideal test in any given circumstance is governed by sample size and by the preference criterion believed to underlie any manipulation.

## III. Application to Ballot Order in Texas and West Virginia Primary Elections.

The main application to which we apply these tests involves the ordering of candidates on primary election ballots in two states, Texas and West Virginia. In these elections, candidate order is to be randomized at the county level by law, generating the repeated orderings needed to conduct our empirical tests. Applying our tests to this data will indicate whether this law was followed uniformly across all counties in each state, or whether the orderings were manipulated in some counties in order to advantage or disadvantage certain candidates. The ballot order effect implies that candidates gain from being listed higher on the ballot, and in our experience this fact is widely understood at the local level.

This application has academic value, as it reveals a variety of departures from randomness, which some methods handle better than others. It has practical relevance as well. County-level randomization is also used in Florida's primary elections, and a broadly similar system in New Hampshire's general election, to which our methods would apply directly. All four states have been the subject of recent ballot order litigation; New Hampshire's system grew out of such litigation (Chaisson, 2018). By our count, some sort of randomization procedure is used in at least some elections in twenty-one states (see Miller, 2010, and Krosnick, Miller, and Tichy, 2004).

Basic Analysis. For 2018 and 2020, we obtained county-level ballot orders online from Texas' and West Virginia's Secretary of State. A total of 41 statewide contests were held in these states in those years, 17 of which have two candidates and utilize the binomial test of equal proportions, with the remainder using the tests analyzed here. For these contests, we utilize all four tests (at least when $N/K! \geq 2$), enabling us to compare their operational performance.

The technical features of these contests vary considerably. Texas has 254 counties, West Virginia just 55, while the number of candidates ranges from 2 to 17. Preferences should also be stronger in high-profile contests for president, senator, and governor than in (say) judicial races further down the ballot. As we will see, there are institutional differences as well.

The results are presented in Table 5, broken down by state, year, and contest. Each non-empty cell has four rows, each containing p-values. The top row pertains to the Equality of Permutations Test–which is equivalent to the binomial test of equal proportions when K=2, and is so indicated by italics. When $K \geq 3$, the second row presents p-values from the Cascading Chi-Squared Test, which can be compared to the respective non-italicized values in the row above it. The third and fourth rows contain the remaining tests, the Rank Test and the Max LC test, which again can be compared to those above them within the same cell. Thus, every vertical column of unitalicized numbers within a cell refers to multiple estimates for a single race (for the West Virginia Supreme Court in 2020, the three vertical columns contain four estimates for each of three races). Empty cells represent uncontested or non-existent races.

Turning first to Texas, on the left side of the table, two contests have a straight run of .00 p-values: the 2018 Democratic primary for U.S. Senate, in which popular candidate Beto O'Rourke was listed first 39% more often than expected, and the 2020 Democratic primary for Railroad Commissioner, in which eventual third-place finisher Kelly Stone was listed first 54% more often than expected. This is a substantial degree of manipulation that could easily sway a close contest (which these were not). In five other races, both Democratic and Republican, the p-values are "suggestive," hovering in the vicinity of .10. In the remaining contests, the p-values–including those from the binomial test–are widely distributed across the unit interval, as would be expected if candidate order is randomly determined in all counties.

Across the 28 Texas contests analyzed, random chance should generate three for which $p \leq .1$ and five or six for which $p \leq .2$. Depending on which test is considered, the actual number of contests meeting those criteria exceeds the expected number by about four. Overall, then, ballot order manipulation is occasionally present in recent Texas primary elections, but is not widespread;

when it occurs, sometimes several counties are involved (as above) and sometimes just a few (generating the small cluster of p-values near .10). This is a lower rate of manipulation than in Grant's (2017) study of the 2014 Texas primaries. The difference may stem from a new state law requiring county chairs to report ballot order to Texas' Secretary of State, who then publishes them on the internet. This sunlight may serve as a disinfectant.

In West Virginia the findings are quite different. The evidence for ballot order manipulation is both weaker and more prevalent. Small p-values such as .00 or .01 are not to be found, because N is only 55, but seven of eleven partisan races have p-values below .20, far greater than the two that would be expected from random chance. (This claim focuses on the more robust tests, for reasons discussed shortly.) Ballot order manipulation is reasonably common in West Virginia, though it is hard to pinpoint the amount that occurs in any given race. Only in judicial elections, which are nonpartisan (but use the same randomization scheme), are the p-values high. In Texas, where judicial elections are partisan, low p-values are common.

Preferences and Power. In Texas, a comparison of tests yields indeterminate results. The p-values are reasonably similar across all four tests and are uniformly low when there is strong evidence of manipulation. Most candidates benefitting from manipulation are popular incumbents or high-profile challengers, which implies unidirectional preferences, yet this does not hold so strongly that the Rank and Max LC Tests dominate.

Circumstances are different in West Virginia, where the Cascading Chi-Squared and Equality of Permutations Tests have an edge in numerous contests. This reflects bidirectional preferences. In the 2020 Democratic primary for U.S. Senator, for instance, two candidates were each listed first and last a disproportionate number of times, with the remaining candidate disproportionately listed in the middle position. As a result, each candidate's mean rank was almost exactly 2.0. The Rank and Max LC Tests, powerless to detect these offsetting preferences, returned large p-values; the other, more flexible tests easily detected this manipulation.

Institutional factors explain why bidirectional preferences are common in West Virginia but not in Texas. In Texas, the county chair of each party has the sole responsibility for conducting that primary's orderings, so all candidates are ordered by a member of their own party. In West Virginia, all orderings are conducted by the county clerk, who is elected in a partisan contest. This means that the same set of candidates is ordered in some counties by a Republican and in other counties by a Democrat. It is not surprising that their preferences are reversed.[8]

Power is enhanced not only by knowing the type of preferences that govern manipulation, but also by knowing the preference criterion itself, which allows targeted tests to be employed. For example, seven candidates entered Texas' 2020 Republican presidential primary, lessening untargeted tests' power. Here, however, one can assume that manipulation would favor the incumbent, Donald Trump, above all challengers (between whom no preferences existed). We assumed this preference criterion in applying GPG's LC and Rank Compatibility Tests to this data. In contrast to the suggestive p-values of the untargeted tests (.02, .09, and .14), both targeted tests rejected the null in favor of this alternative with p-values of .00.

[8] This being said, bidirectional preferences aren't wholly absent in Texas. GPG identify one contest from 2014, the Republican primary for Comptroller, that also displayed such preferences. That year, strife between the establishment and Tea Party factions of the party was especially prevalent; things had largely smoothed over by the time of the 2018 and 2020 races analyzed here.

**Section IV. Supplementary Applications: Powerball and *American Idol*.**

The variety of outcomes observed in the ballot order context demonstrated the value of having a battery of tests at one's disposal. The two applications presented in this section further develop this theme, while highlighting the range of questions these tests can address and the implementation issues that can arise in doing so.

Powerball. Powerball, the best known lottery in the U.S., is conducted semi-weekly by the Multi-State Lottery Association, with tickets sold in 45 states and some U.S. territories. Held since 1992, the lottery selects five numbered "white balls" plus a red numbered "Powerball." Each is drawn physically and sequentially using mechanical devices that resemble popcorn poppers. The Wisconsin Lottery had posted over fifteen years' of winning numbers in the order that they were drawn, a total of 1,596 drawings. We use this data for our analysis.

To test the null hypothesis that the white balls are drawn randomly, one could analyze the *frequency* with which they are chosen. However, a ball would be drawn "too much" only if it had a mechanical flaw that caused it to edge out other balls for selection. If so, it should tend to be drawn earlier in the sequence as well. With sufficient data, a test of the randomness of the *orderings* should outweigh one that merely examines frequencies, since it incorporates information on the sequence in which the balls are drawn.

Any deviations from randomness should be unidirectional, just as in our simulations, where smaller numbers were more likely to be selected for the next position in the sequence. However, the tests best suited to these "preferences," the Rank and Max LC Tests, cannot be applied here, as only a subset of five white balls are ordered each week. The methods in this paper are not designed for *partially blocked* data such as this, only *fully blocked* data in which every item is ordered. How can we test the randomness of incomplete orderings like these?

The solution is to adapt the Cascading Chi-Squared Test so that, instead of proceeding by *item*, examining at the positions of the 1 ball across all orderings and so on, it proceeds by *position*. Initially, we count the frequency each ball is drawn first in the set of orderings and calculate the appropriate chi-squared statistic. Then we count the frequency each ball is drawn second and calculate a second chi-squared statistic in the usual way, adjusting the expected frequency of each ball to account for those that have already been drawn for first position. We continue this process for all five positions that are drawn and sum these five statistics. As the assumption of independence is not satisfied for every position after the first, p-values are calculated using Monte Carlo simulation, though deviations from the standard $\chi^2$ distribution are small in this instance.[9]

The results are placed in the first column of Table 6, first for three sub-periods with increasing values of K, and then for the full sample, handling the variation in K by adjusting the expected number of occurrences of each ball accordingly. The p-values are generally large; there is no evidence of "manipulation." The results are similar for the counts, in the second column of the table, and there is no clear sign either test is superior. This is not surprising, as deviations from randomness are unlikely to occur in a highly visible, mechanized process in which the incentives run the other way.

[9] For fully blocked data, these statistics' sum has a distribution that closely resembles the chi-squared, with an expected value of $K(K - H_K)$, where $H_K$ is the Kth harmonic number; the test is slightly undersized. This problem is smaller for partially blocked data, as the non-independence problem is "weaker."

In a plot twist, however, mechanical order manipulation *does* appear elsewhere in our data. It explains Texas' 2020 Democratic primary for Railroad Commissioner, mentioned above, which gave third place finisher Kelly Stone top billing. In a replay of the $\delta = 0.8$ table in Appendix B, Stone was placed in first position 98 times, second position 70 times, third position 54 times, and last position 32 times. The (rough) reverse was true for the two challengers who beat her. At each step in the sequence, Stone was favored over these other candidates–the hallmark of the mechanical process posited for Powerball and utilized in our simulations.

Again, institutional factors can explain how this occurs. Ballot order is often determined by drawing wads of paper from a bowl. If these were not identically shaped and sized, some would have an advantage for selection and would tend to be drawn earlier in the sequence. While counties' ballot orders are often drawn locally, many Texas counties, rural and heavily Republican, have no local Democratic contests. For statewide races, these counties' ballot orders are often drawn by the state party in Austin.[10] There, the same wads would be used repeatedly for a given contest, yielding the cross-tabulation we observe.

*American Idol*. The popular television show *American Idol* ran on the Fox television network from 2002-2016.[11] Each season pares down an initial set of roughly twelve finalists to a single winner, generally by eliminating one finalist each week. Page and Page (2010) have shown that contestants' success is influenced by the sequence in which they perform on the program, with earlier-performing

---

[10] Glen Maxey, staff member of the Texas Democratic Party, in personal communication.

[11] The show has since moved to a different network. As this other network could use a different procedure for ordering the contestants, its seasons of the show are not analyzed.

singers receiving fewer votes from the television audience and exiting at higher rates.

There is little direct evidence on how this sequence is determined each week. Though not literally randomly determined, the process may still be as good as random for practical purposes:

> Producers decide the singing order except for the finale, which is a singer's choice after a coin flip. They vary the order each week to be fair, but also try to arrange singers and their songs to make the most entertaining show, executive producer Ken Warwick says. "It's worked out with two things in mind: where the kid (performed) last week, and if it's a slow, 'dirgey' ballad, I try not to open with that," he says. ("'Idol' Singers Who Go First May Not Last," *Norwich Bulletin*, Apr. 21, 2008).

Broadly speaking, then, there are three possibilities:

- The orderings are as good as randomly determined.
- Certain performers are systematically placed earlier (later) in the program, perhaps because they regularly perform upbeat (downbeat) songs.
- The orderings are conducted so as to even out imbalances over time, that is, to "correct" random inequities in earlier weeks, out of concern for fairness.

The possibility of fairness-based manipulation stands traditional hypothesis testing on its head, as this manipulation should raise p-values, not lower them. We sidestep this problem by examining the p-values from multiple seasons of the show. If these are widely spread across the unit interval, the first possibility is most credible. If they cluster below .5, the second possibility is most credible. If they cluster above .5, so that there is less variation than is implied by random chance, the third possibility is most credible.

Again, however, the data are not fully blocked–indeed, they are *unblocked*, as the number of contestants falls each week. Again we must adapt one of our tests for randomness. The best candidate is the Max LC Test. In the strictly untargeted version of this test, the score vector can only take those values that are optimal under some preference ranking, assuming that the strength of these preferences is not known. To adapt this test to this data, let the elimination of any one contestant leave unchanged the preferences over those who remain. If Contestant 1 ≻ Contestant 2 ≻ Contestant 3 ≻ Contestant 4, and Contestant 3 is eliminated that week, then the next week's preferences are Contestant 1 ≻ Contestant 2 ≻ Contestant 4. Then, given any initial preference ordering, one can calculate the concordance L in equation **(1)** for any subset of items, by modifying K and the score vector accordingly. Doing so for each week of the season, one can calculate the mean concordance across the season for any and all initial preference orderings. The modified Max LC statistic is associated with the hypothesized preferences that yield the largest mean concordance. The p-value associated with this statistic is calculated through Monte Carlo simulation.

This is simple enough in theory, but in practice the number of possible preference rankings is large. Allowing indifference between contestants, that is, ties in these rankings, there are 1.6 billion possible rankings when K = 11. As this application is merely illustrative, we simplify matters by limiting K to 10 and excluding preference criteria that allow indifference between contestants. Accordingly, we analyze the last nine weeks of the ten seasons of *American Idol*, listed in Table 7, that eliminate exactly one contestant per week over that period (or close enough to it that any deviations can be finessed, as explained in the note to the table).

The p-values for these ten seasons, reported in this table, are widely disbursed across the unit interval, with a mean of .46, and the Kolmogorov-Smirnov test for uniformity returns a p-value of 0.18. This supports the first possibility listed above. In the Fox years of *American Idol*, contestant orderings were as good as randomly conducted. In their analysis of order effects on success probabilities in *American Idol*, Page and Page (2010) did not detect any bias in their estimates arising from non-random sequencing. Our finding supports this conclusion and extends it.

## V. Discussion and Conclusion.

The "ordering problem" is surprisingly rich. It appears in a variety of contexts, in which deviations from randomness could be anti-social (self-interested ballot order determination), pro-social (fairness concerns in *American Idol*), or asocial (mechanical flaws in Powerball). Preferences can be strong, weak, or non-existent; unidirectional, bidirectional, or otherwise. Outcomes are meaningfully affected by small institutional details.

Such varied circumstances impact the technical aspects of testing for the randomness of a set of repeated orderings. Statistical power ranges from weak to strong, depending on the number of items being ordered, the number of orderings being conducted, and the strength of preferences, all of which varied widely in our applications. The structure of the data varied as well: election ballots ordered all items, Powerball drawings a constant subset of items, *American Idol* a decreasing subset of items. Similarly, in some instances, the preference criterion could be anticipated in advance; in others only the type of preferences; in still others, neither were known.

This variation means that there is no single best test for the randomness of repeated orderings. When there is an adequate basis to specify a preference criterion in advance, targeted tests are preferred; otherwise, untargeted tests must be used. When preferences are unidirectional and power is limited, the Rank Test and Max LC Test are preferred. Otherwise, the Cascading Chi-Squared and Equality of Permutation Tests are likely to prevail. When the data are not fully blocked, it may be necessary to adapt the Cascading Chi-Squared or Max LC tests accordingly, sometimes increasing computational costs in consequence.

In other words, to be fully equipped for empirical work in this area, researchers must understand the landscape of the problem and have a battery of tests at their disposal. Accordingly, this paper has considered six such tests–three old, three new–and shown how they can be adapted to suit unusual data structures. We hope this toolkit will equip researchers to examine the randomness of repeated orderings in the wide variety of places in which they occur in everyday life.

Table 1. Power Simulations, Rank Test (power at $\alpha = .05$ for $\delta = .95, .9, .8$, and .7, in that order).

| K↓ N→ | 50 | 125 | 250 | 1000 |
|---|---|---|---|---|
| 3 | .07, .12, .33, .69 | .07, .17, .63, .97 | .10, .33, .93, 1 | .31, .89, 1 |
| 4 | .07, .16, .60, .96 | .10, .35, .96, 1 | .18, .66, 1 | .64, 1 |
| 5 | .09, .28, .89, 1 | .16, .62, 1 | .34, .94, 1 | .92, 1 |
| 6 | .14, .46, .99, 1 | .26, .88, 1 | .54, 1 | .99, 1 |
| 10 | .42, .99, 1 | .87, 1 | 1 | 1 |
| 15 | .93, 1 | 1 | | |
| 20 | 1 | | | |

Table 2. Power Simulations, Freely Untargeted, Strictly Untargeted, and Targeted Max LC Tests (power at $\alpha = .05$ for $\delta = .95, .9, .8$, and .7, in that order).

Freely Untargeted

| K↓ N→ | 50 | 125 | 250 | 1000 |
|---|---|---|---|---|
| 3 | .06, .09, .29, .63 | .08, .18, .67, .96 | .11, .32, .93, 1 | .31, .90, 1 |
| 4 | .07, .16, .61, .96 | .11, .36, .96, 1 | .18, .67, 1 | .65, 1 |
| 5 | .09, .27, .89, 1 | .17, .63, 1 | .33, .93, 1 | .93, 1 |
| 6 | .12, .44, .99, 1 | .26, .87, 1 | .54, 1 | 1 |
| 10 | .41, .99, 1 | .87, 1 | 1 | |
| 15 | .93, 1 | 1 | | |
| 20 | 1 | | | |

Strictly Untargeted

| K↓ N→ | 50 | 125 | 250 | 1000 |
|---|---|---|---|---|
| 3 | .06, .11, .30, .64 | .08, .19, .66, .98 | .12, .35, .93, 1 | .32, .90, 1 |
| 4 | .08, .17, .64, .97 | .11, .37, .97, 1 | .20, .68, 1 | .67, 1 |
| 5 | .10, .30, .91, 1 | .18, .68, 1 | .33, .95, 1 | .93, 1 |
| 6 | .13, .46, .99, 1 | .29, .90, 1 | .56, 1 | |

Targeted

| K↓ N→ | 50 | 125 | 250 | 1000 |
|---|---|---|---|---|
| 3 | .11, .20, .50, .83 | .14, .33, .83, .99 | .22, .55, .98, 1 | .54, .97, 1 |
| 4 | .15, .35, .86, 1 | .25, .64, 1 | .40, .89, 1 | .87, 1 |
| 5 | .23, .58, .99, 1 | .42, .90, 1 | .65, .99, 1 | .99, 1 |
| 6 | .33, .80, 1 | .60, .99, 1 | .86, 1 | 1 |
| 10 | .86, 1 | 1 | | |
| 15 | 1 | | | |
| 20 | | | | |

Note: the strictly untargeted test was very slow to run for $N \geq 10$ and so was omitted. Power for omitted entries is implicitly one, as simulations determined it to be one for less favorable values of N, K, and/or $\delta$.

Table 3. Power Simulations, Cascading Chi-Squared Test (power at $\alpha = .05$ for $\delta = .95, .9, .8$, and .7, in that order).

| K↓ N→ | 50 | 125 | 250 | 1000 |
|---|---|---|---|---|
| 3 | .06, .09, .25, .57 | .07, .16, .57, .95 | .09, .28, .89, 1 | .27, .85, 1 |
| 4 | .07, .13, .47, .91 | .09, .28, .91, 1 | .14, .55, 1 | .53, 1 |
| 5 | .07, .18, .75, 1 | .12, .46, 1 | .22, .83, 1 | .81, 1 |
| 6 | .08, .27, .93, 1 | .16, .69, 1 | .35, .98, 1 | .97, 1 |
| 10 | .20, .82, 1 | .55, 1 | .93, 1 | 1 |
| 15 | .49, 1 | .98, 1 | 1 | |
| 20 | .87, 1 | 1 | | |

Table 4. Simulation Results, Untargeted and Targeted Equality of Permutation Tests (power at $\alpha = .05$ for $\delta = .95, .9, .8$, and .7, in that order).

Untargeted

| K↓ N→ | 50 | 125 | 250 | 1000 |
|---|---|---|---|---|
| 3 | .05, .07, .21, .51 | .07, .13, .51, .93 | .08, .24, .85, 1 | .22, .80, 1, 1 |
| 4 | *.06, .09, .28, .71* | .07, .16, .71, 1 | .10, .32, .98, 1 | .31, .97, 1, 1 |
| 5 | ---- | ---- | *.10, .33, .99, 1* | .30, .99, 1, 1 |

Targeted

| K↓ N→ | 50 | 125 | 250 | 1000 |
|---|---|---|---|---|
| 3 | .05, .09, .23, .49 | .08, .17, .49, .85 | .11, .26, .74, .98 | .29, .73, 1, 1 |
| 4 | *.04, .06, .18, .43* | .08, .17, .50, .87 | .10, .24, .75, .98 | .27, .70, 1, 1 |
| 5 | ---- | ---- | *.05, .14, .50, .92* | .19, .52, .98, 1 |

Note: Results in italics satisfy $2K! \leq N < 5K!$.

Table 5. P-values from Randomization Tests, Texas and West Virginia Primaries (in vertical order within each cell, p-values from the Equality of Permutations Test, the Cascading Chi-Squared Test, the Rank Test, and the freely untargeted Max LC Test).

<table>
<tr><th></th><th colspan="4">TEXAS</th><th colspan="4">WEST VIRGINIA</th></tr>
<tr><th></th><th colspan="2">2018</th><th colspan="2">2020</th><th colspan="2">2018</th><th colspan="2">2020</th></tr>
<tr><th>Office</th><th>Democratic</th><th>Republican</th><th>Democratic</th><th>Republican</th><th>Democratic</th><th>Republican</th><th>Democratic</th><th>Republican</th></tr>
<tr><td>President</td><td>----</td><td>----</td><td>--<br>.81<br>.33<br>.38</td><td>--<br>.14<br>.09<br>.02</td><td>----</td><td>----</td><td>--<br>.15<br>.24<br>.20</td><td>--<br>.15<br>.54<br>.56</td></tr>
<tr><td>U.S. Senate</td><td>.00<br>.00<br>.00<br>.00</td><td>.06<br>.34<br>.08<br>.10</td><td>--<br>.67<br>.54<br>.54</td><td>.48<br>.55<br>.44<br>.47</td><td>1.00<br>--</td><td>--<br>.14<br>.09<br>.09</td><td>.06<br>.04<br>.54<br>.55</td><td>.17<br>.15<br>.31<br>.33</td></tr>
<tr><td>Governor</td><td>--<br>.70<br>.87<br>.88</td><td>.48<br>.66<br>.88<br>.92</td><td>----</td><td>----</td><td>----</td><td>----</td><td>--<br>.17<br>.14<br>.17</td><td>--<br>.03<br>.04<br>.03</td></tr>
<tr><td>Agriculture Comm.</td><td>----</td><td>.35<br>.30<br>.09<br>.11</td><td>----</td><td>----</td><td>----</td><td>----</td><td>.31<br>.25<br>.22<br>.21</td><td>1.00<br>--</td></tr>
<tr><td>Other State Offices</td><td>.85,.15,.23,.75<br>--</td><td>.75, .12, .04<br>.13<br>.29<br>.29</td><td>.00<br>.00<br>.00<br>.00</td><td>.75<br>--</td><td>----</td><td>----</td><td>.59<br>--</td><td>----</td></tr>
<tr><td>Supreme Court</td><td>----</td><td>----</td><td>.95,.75,.49,.49</td><td>----</td><td colspan="2">----</td><td colspan="2">.74, .81, .07<br>.58, .92, .32<br>.43, .41, .73<br>.47, .41, .74</td></tr>
<tr><td>Court of Criminal Appeals</td><td>----</td><td>.09, .05<br>.12<br>.01<br>.02</td><td>.57, .11<br>.10<br>.08<br>.09</td><td>.19<br>--</td><td colspan="2">----</td><td colspan="2">----</td></tr>
</table>

Note: Each line contains p-values for all contests fitting the description for that cell. In West Virginia, Supreme Court positions are non-partisan races placed in the same ballot order on both parties' primary ballots. Supreme Court positions contested in Texas in 2018 were were Places 2, 4, and 6; in 2020, the Chief Justice and Places 6, 7, and 8; Court of Criminal Appeals positions contested in 2018 were the Presiding Judge and Place 8; in 2020 were Places 3 and 4. "Other State Offices" includes Attorney General in West Virginia and, in Texas, Lt. Governor, Comptroller, and Land Commissioner in 2018 and Railroad Commissioner in both 2018 and 2020. Italicized numbers involve binomial tests in two-candidate races. When the Equality of Permutations test is not presented, but other multi-candidate tests are presented, it is because the requirement $N/K! \geq 2$ was not met.

Table 6. Randomization Tests of Powerball Drawings ($\chi^2$ test statistic, with p-values in parentheses).

| Time Frame | Modified Cascading $\chi^2$ Test | Equality of Proportions Test |
|---|---|---|
| Aug. 31, 2005 - Jan. 03, 2009<br>(K = 55, N = 350) | 245.1<br>(0.85) | 46.2<br>(0.76) |
| Jan. 07, 2009 - Oct. 3, 2015<br>(K = 59, N = 704) | 258.2<br>(0.90) | 37.7<br>(0.98) |
| Oct. 7, 2015 - June 1, 2019<br>(K = 69, N = 382) | 364.9<br>(0.16) | 75.7<br>(0.24) |
| Feb. 18, 2004 - June 1, 2019<br>(K = 69*, N = 1,596) | 307.9<br>(0.93) | 63.4<br>(0.63) |

* K = 53 from Feb. 18, 2004 - Aug. 27, 2005, and then grew as listed in the table.

Table 7. Results, American Idol, Modified Max LC Test.

| | p-value(s) |
|---|---|
| Season 2 | 0.37 |
| Season 3 | 0.15 |
| Season 4 | 0.81 |
| Season 5 | 0.38 |
| Season 7 | 0.93 |
| Season 10 | 0.47 / 0.49 |
| Season 11 | 0.04 |
| Season 12 | 0.51 |
| Season 13 | 0.61 |
| Season 14 | 0.30 / 0.37 |

Note: Season 2 removed Corey Clark, who was disqualified in an early round, from all rankings. In Seasons 10 and 14, there were 11 people in the first round analyzed, two of which were eliminated before the consequent round. For these seasons, the test was run twice, first retaining one, then the other, of these two contestants. In several seasons, there were earlier rounds with additional contestants; only the last nine rounds were analyzed.

## Appendix A

Define $\alpha_j$ as item j's sum of ranks, that is, the number of times it is ranked first, plus two times the number of times it is ranked second, and so on; i.e., $\Sigma(pC_{j,p})$, where p indexes position or rank. Then, the program to be solved can be written as follows:

$$Lagrangian = \sum_{i=1}^{K} \alpha_i s_i - \lambda_1 \sum_{i=1}^{K} s_i - \lambda_2 (\sum_{i=1}^{K} s_i^2 - 1) \quad \textbf{(6)}$$

where $\lambda_1$ and $\lambda_2$ are Lagrange Multipliers associated with the constraints that the score vector s have a mean of zero and a 2-norm of one.

Taking the derivative with respect to $s_j$ yields the following first order condition:

$$s_j^* = (\alpha_j - \lambda_1)/2\lambda_2 \quad \forall j \quad \textbf{(7)}$$

Summing the $s_j^*$ and setting equal to zero yields:

$$\sum_{j=1}^{K} s_j^* = \sum_{j=1}^{K} (\alpha_j - \lambda_1)/2\lambda_2 = 0 \quad \Rightarrow \quad \lambda_1 = \frac{1}{K} \sum_{j=1}^{K} \alpha_j = \bar{\alpha} \quad \textbf{(8)}$$

As $\bar{\alpha} = N(K+1)/2$, for all j, $s_{j*}$ is proportional to $\alpha_j$ - N(K+1)/2, as claimed in the text.

**Appendix B**

Table B1. Relative Frequencies of Positions by Item, K=4, for the δ Values Used in Tables 1-4.

| δ = 0.95 | First Position | Second Position | Third Position | Fourth Position |
|---|---|---|---|---|
| First Item | 0.276 | 0.257 | 0.238 | 0.229 |
| Second Item | 0.260 | 0.253 | 0.249 | 0.237 |
| Third Item | 0.245 | 0.251 | 0.251 | 0.253 |
| Fourth Item | 0.219 | 0.239 | 0.262 | 0.280 |

| δ = 0.90 | First Position | Second Position | Third Position | Fourth Position |
|---|---|---|---|---|
| First Item | 0.295 | 0.258 | 0.229 | 0.218 |
| Second Item | 0.268 | 0.255 | 0.255 | 0.222 |
| Third Item | 0.244 | 0.262 | 0.251 | 0.244 |
| Fourth Item | 0.193 | 0.225 | 0.265 | 0.317 |

| δ = 0.80 | First Position | Second Position | Third Position | Fourth Position |
|---|---|---|---|---|
| First Item | 0.344 | 0.267 | 0.213 | 0.176 |
| Second Item | 0.288 | 0.271 | 0.238 | 0.204 |
| Third Item | 0.228 | 0.257 | 0.268 | 0.247 |
| Fourth Item | 0.140 | 0.208 | 0.282 | 0.374 |

| δ = 0.70 | First Position | Second Position | Third Position | Fourth Position |
|---|---|---|---|---|
| First Item | 0.394 | 0.279 | 0.196 | 0.131 |
| Second Item | 0.311 | 0.282 | 0.234 | 0.173 |
| Third Item | 0.202 | 0.270 | 0.282 | 0.246 |
| Fourth Item | 0.093 | 0.169 | 0.289 | 0.449 |

Note: These relative frequencies are calculated from 10,000 simulations for each value of δ.

## Appendix C

To generalize the simulations in Section II, we give numerical values $v_i$ to each item, such that A=1, B=2, and so on. The probability that each item is selected is then set to be a function of the absolute difference between these values and the agent's analogous $v_a$ value, thusly:

$$
\begin{aligned}
&P(selection) \propto \delta^{|v_a - v_i|} \\
&\textit{where} \\
&\textit{Unidirectional}: \; v_a = 0 \\
&\textit{Bidirectional}: \; P(v_a{=}0) = P(v_a{=}K{+}1) = .5 \\
&\textit{Multidirectional}: \; v_a \sim U(0,K{+}1)
\end{aligned}
\qquad \textbf{(9)}
$$

When $v_a = 0$, this process devolves to that used in Section II; thus this mechanism can be considered a generalization of that process to incorporate other preference types.

Select simulation results are presented in Table C1. As in the text, unidirectional preferences are best detected by the Rank and Max LC Tests. The other two tests are more robust and may, but need not, dominate for other preference structures. Under those circumstances, the fully general Equality of Permutations Test is preferred when the condition $N \geq 2K!$ can be met; otherwise, the Cascading Chi-Squared Test must suffice.

For multidirectional preferences, the p-values are similar across tests. This is not inherent to this preference structure, but is instead an artifact of the relative variation in agents and candidates specified in equation **(9)**. Decreasing the spread of agents relative to candidates, say by replacing U(0,K+1) with U(1,K), would make the Rank and Max LC Tests dominate; increasing this spread would favor the other two tests. (While additional simulations affirm this claim, it can also be justified intuitively: in the limit, as the spread of agents increases, preferences become bidirectional.) Therefore, the fact that preferences are multidirectional does not determine the optimal test, and the dominance of the Rank and Max LC Tests need not indicate unidirectional preferences.

Table C1. Simulation Results for Select Values of K and δ.

| K = 4, δ = 0.7 | | | | | | | | | |
|---|---|---|---|---|---|---|---|---|---|
| | N = 50 | | | N = 250 | | | N = 1000 | | |
| | Unidirect. | Bidirect. | Multidirect. | Unidirect. | Bidirect. | Multidirect. | Unidirect. | Bidirect. | Multidirect. |
| Rank Test | .96 | .05 | .08 | 1 | .05 | .27 | 1 | .06 | .84 |
| Max LC Test | .96 | .05 | .08 | 1 | .05 | .26 | 1 | .05 | .84 |
| Cascading $\chi^2$ Test | .91 | .06 | .08 | 1 | .12 | .23 | 1 | .43 | .82 |
| Equality of Permutations Test | .71 | .09 | .08 | 1 | .32 | .26 | 1 | .96 | .91 |
| K = 7, δ = 0.8 | | | | | | | | | |
| | N = 50 | | | N = 250 | | | N = 1000 | | |
| | Unidirect. | Bidirect. | Multidirect. | Unidirect. | Bidirect. | Multidirect. | Unidirect. | Bidirect. | Multidirect. |
| Rank Test | 1 | .05 | .14 | 1 | .06 | .64 | 1 | .05 | 1 |
| Max LC Test | 1 | .05 | .12 | 1 | .06 | .65 | 1 | .05 | 1 |
| Cascading $\chi^2$ Test | 1 | .09 | .11 | 1 | .30 | .53 | 1 | 1 | 1 |
| Equality of Permutations Test | ---- | ---- | ---- | ---- | ---- | ---- | ---- | ---- | ---- |